\documentclass[5p,times, number]{elsarticle}
\usepackage{flushend}
\usepackage{lipsum}
\usepackage{mathtools}
\usepackage{cuted}
\usepackage{float}
\usepackage{graphicx}
\graphicspath{ {./Images/} }
\usepackage{cite}
\usepackage[super]{nth}
\usepackage{amsmath,amssymb}
\DeclareMathOperator{\EX}{\mathbb{E}}
\usepackage{pdfpages}

\begin{document}

\begin{minipage}{\textwidth}
\includepdf[scale=0.99]{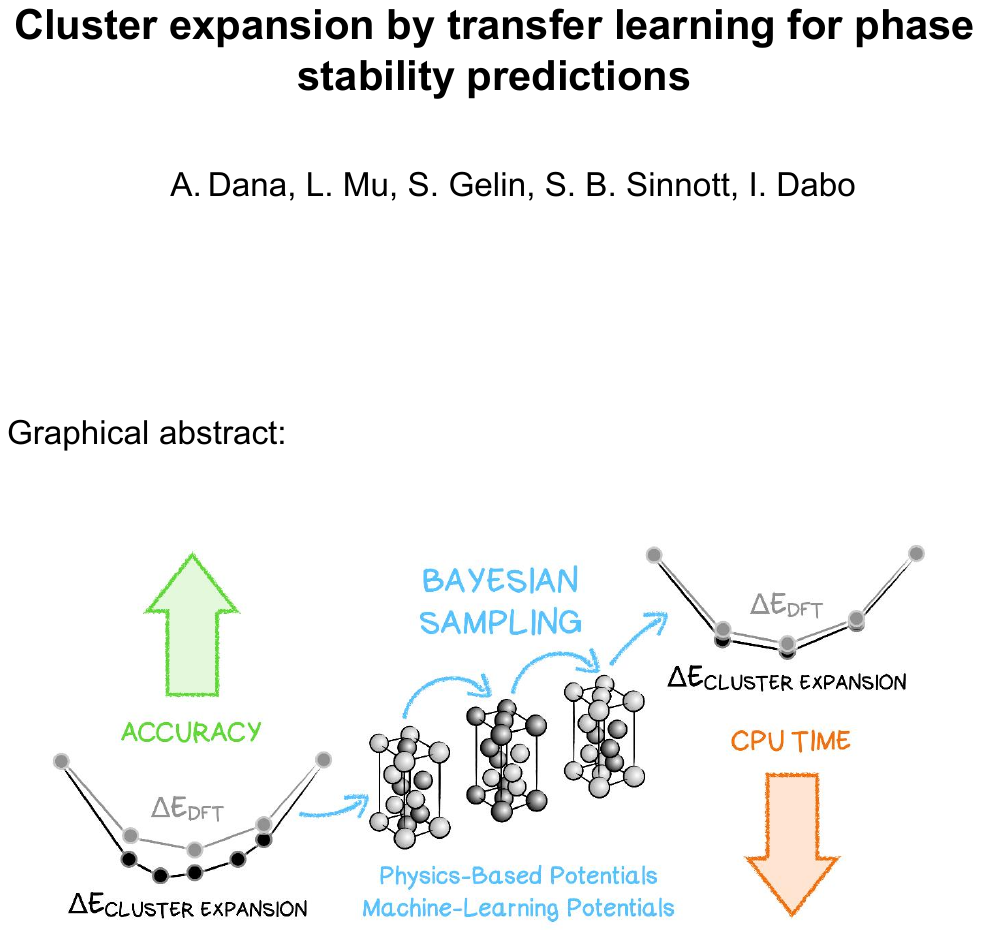}
\end{minipage}

\begin{frontmatter}

\title{\Large {\bf Cluster expansion by transfer learning for phase stability predictions}}

\author[MatSE,MRI]{A.~Dana\corref{cor1}}
\ead{a.dana@psu.edu}
\author[MatSE]{L.~Mu}
\author[MatSE,MRI]{S.~Gelin}
\author[MatSE,MRI,Chemistry,ICDS]{S.~B.~Sinnott}
\author[MatSE,MRI,Physics]{I.~Dabo\corref{cor1}}
\ead{dabo@psu.edu}
\cortext[cor1]{Corresponding authors}
\address[MatSE]{Department of Materials Science and Engineering, The Pennsylvania State University, University Park, PA 16802, USA.}
\address[MRI]{Materials Research Institute, The Pennsylvania State University, University Park, PA 16802, USA.}
\address[Chemistry]{Department of Chemistry, The Pennsylvania State University, University Park, PA 16802, USA.}
\address[ICDS]{Institute for Computational and Data Sciences, The Pennsylvania State University, University Park, PA 16802, USA.}
\address[Physics]{Department of Physics, The Pennsylvania State University, University Park, PA 16802, USA.}

\begin{abstract}
  Recent progress towards universal machine-learned interatomic potentials holds considerable promise for materials discovery. Yet the accuracy of these potentials for predicting phase stability may still be limited. In contrast, cluster expansions provide accurate phase stability predictions but are computationally demanding to parameterize from first principles, especially for structures of low dimension or with a large number of components, such as interfaces or multimetal catalysts. We overcome this trade-off {\it via} transfer learning. Using Bayesian inference, we incorporate prior statistical knowledge from machine-learned and physics-based potentials, enabling us to sample the most informative configurations and to efficiently fit first-principles cluster expansions. This algorithm is tested on Pt:Ni, showing robust convergence of the mixing energies as a function of sample size with reduced statistical fluctuations.
\end{abstract}

\begin{keyword}
\it{Bayesian sampling; density-functional theory; graph neural networks; reactive potentials; embedded-atom potentials}
\end{keyword}

\end{frontmatter}

\section{Introduction}

\begin{figure*}
    \centering 
    \includegraphics[width=\textwidth]{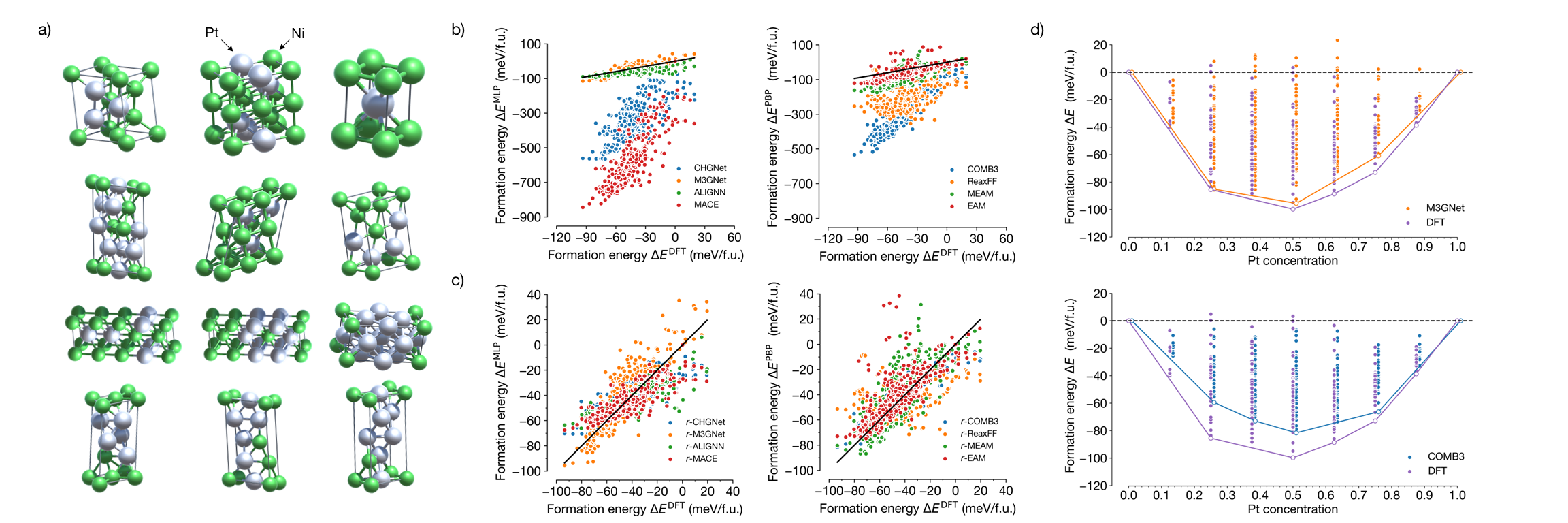}
    \caption{Accuracy of machine-learned and physics-based (reactive and embedded-atom) potentials in reproducing DFT energies [within the Perdew--Burke--Ernzerhof (PBE) generalized-gradient approximation]. (a) 12 representative supercells (out of a dataset of 413 symmetrically inequivalent structures) for Pt:Ni. (b) Parity plots of mixing energies for machine-learned and physics-based potentials relative to the DFT reference and (c) parity plots after optimal rescaling of the energies with respect to DFT (cf.~Sec.~\ref{sec:results} for a detailed description of the rescaling method). (The prefix `$r$-' indicates that the potential is rescaled.) (d) Convex hull plots for M3GNet and COMB3 using the scaled energies. (The energy points are slightly shifted with respect to the actual concentrations for ease of comparison.) }
    \label{ml parity}
\end{figure*}

Accurate and efficient predictions of phase stability are critical to materials discovery. While machine-learned potentials can efficiently explore the configurations of a phase, their precision is limited when these configurations are not captured by the training dataset. To quantify this limitation, Fig.~\ref{ml parity} compares the accuracy of select interatomic potentials, including pre-fitted many-body potentials (charge-optimized many-body potential, COMB3 \citep{liang2013classical}; reactive force field, ReaxFF \citep{Shin2016DevelopmentCatalyst}; embedded-atom method, EAM \citep{Foiles1986Embedded-atom-methodAlloys}; modified embedded-atom method, MEAM \citep{Kim2017SecondSystems}) and off-the-shelf machine-learning models (crystal Hamiltonian graph neural network, CHGNet \citep{deng2023chgnet}; graph neural network with three-body interactions, M3GNet \citep{chen2022universal}; atomistic line graph neural network, ALIGNN \citep{choudhary2021atomistic}; message passing multilayer atomic cluster expansion, MACE \citep{Batatia2022mace,Batatia2022Design}) in predicting the stability of the face-centered cubic Pt:Ni binary [Fig.~\ref{ml parity}(a)]. Reactive and embedded-atom physics-based potentials (PBPs) rely on predetermined, physically formulated functions, often tailored to specific chemical compositions; as such, they may inherit some transferability beyond the limits of the training data. In contrast, machine-learned potentials (MLPs) do not typically depend on physical approximations; they utilize highly adaptable analytical formulations to predict potential energies and should generally be restricted to the regions of the configurational space that are covered by the training dataset \citep{eckhoff2023lifelong}. 

As shown in Fig.~\ref{ml parity}(b), for this prototypical bimetallic alloy, MLP and PBP energies can deviate considerably from density-functional theory (DFT) calculations (as MLPs and PBPs may not adequately extrapolate DFT predictions). Although the results for MEAM, EAM, ALIGNN, and M3GNet appear relatively close to the DFT reference for Pt:Ni, discrepancies of up to 40 meV per atom are still observed. While MLPs and PBPs do not yet achieve the precision of DFT models, they still carry important information about the relative energies of the different configurations, as illustrated in Fig.~\ref{ml parity}(c), which presents rescaled formation energies (the calculation of the scaling factor is explained in Sec.~\ref{sec:results}). Convex hull diagrams are also presented in Fig.~\ref{ml parity}(d), demonstrating that MLPs and PBPs capture the ordering of formation energies to a reasonable extent, although they do not reliably predict the convex hull of the stable configurations for the face-centered cubic  phase of Pt:Ni.

We circumvent this limitation by accelerating the parameterization of cluster expansions (CEs), exploiting the latent information contained in MLP and PBP data. CEs evaluate the energy of a lattice by summing energy contributions from finite-size clusters across lattice sites \citep{kadkhodaei2021cluster,nelson2013cluster}. These models have been widely used to study crystalline order and phase stability at reduced computational cost relative to DFT calculations \citep{wolverton1998prediction, seko2006prediction, kolb2005nonmetal,carlsson2023finding, wu2016cluster}, and are useful for predicting free energies \citep{ozolicnvs1998cu, wolverton1998prediction, seko2006prediction, kolb2005nonmetal}, magnetic states \citep{sanchez1989magnetic}, phase transitions \citep{kadkhodaei2021cluster, ozolicnvs1998cu, asta1993theoretical}, and defect stability \citep{van2005vacancies}. The central complication in constructing CEs is to generate a dataset of DFT energies. This constraint is especially problematic for low-dimension systems, as the absence of full translation symmetry implies that a large number of configurations is needed to capture the interatomic interactions along the nonperiodic direction(s) \citep{cao2018use, nelson2013cluster}. Considerable effort has been dedicated to generating cluster expansions that minimize prediction errors for a given training set size. While various machine-learning techniques, including active learning \citep{van2002automating, seko2009cluster, mueller2010exact}, cross validation \citep{van2002automating}, regularization \citep{cockayne2010building,mueller2009bayesian}, and feature selection \citep{cockayne2010building,mueller2009bayesian, nelson2013compressive, tibshirani1996regression}, are commonly employed to circumvent this bottleneck, significant improvements in computational efficiency may be achieved by leveraging statistical correlations extracted from MLPs and PBPs.

In what follows, we present and validate an algorithm to expedite the fitting of cluster expansions by transfer learning. This approach exploits Bayesian inference to extract prior knowledge from MLPs/PBPs, enabling one to identify of the most informative configurations in a given pool \citep{packwood2017bayesian, mueller2009bayesian, todorovic2019bayesian}. We show the efficacy of this method by examining Pt:Ni intermetallics.

\section{Methodology}

\label{sec:methodology}

\subsection{Cluster expansion}

\begin{figure*}
    \centering
    \includegraphics[width=0.85\textwidth]{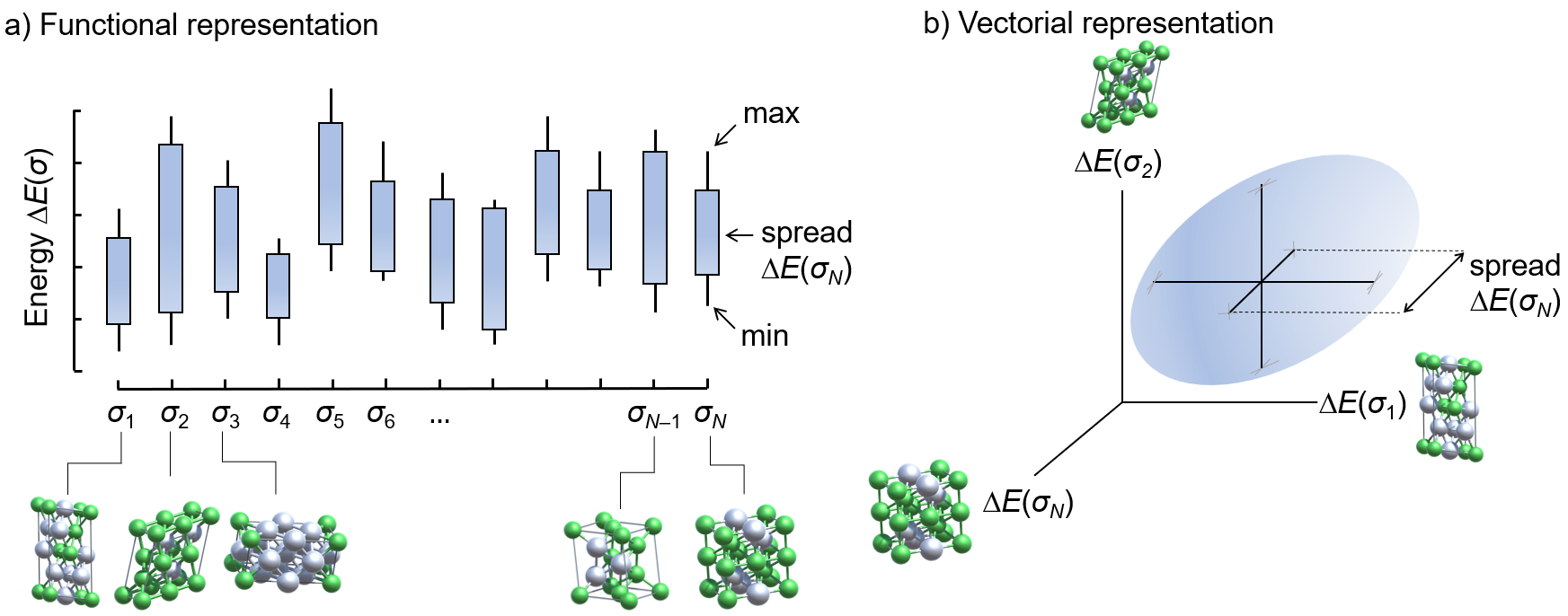}
    \caption{Functional and vectorial representations of the statistical distribution of the energies of $N$ configurations. (a) By calculating the average and spread of the energy of each cluster across selected empirical potentials, one obtains a statistical energy distribution. (b) This distribution can be represented as a Gaussian probability in $N$-dimensional vector space of the configurational energies.}
    \label{gaussian}
\end{figure*}

Within cluster expansions, the formation energy $\Delta E$ of a configuration \({\boldsymbol \sigma}\) of a system is expressed as the sum of energy contributions associated with symmetrically inequivalent clusters that make up that configuration \citep{wu2016cluster,sanchez2016approximate}:
\begin{equation}
\Delta E({\boldsymbol \sigma})=\sum_\alpha{m_{\alpha} J_{\alpha} \pi_\alpha({\boldsymbol \sigma})},
\label{eq:ce}
\end{equation}
where \(\pi_\alpha({\boldsymbol \sigma})=\left\langle\prod_{i}\varphi_{\alpha,i}(\sigma_i)\right\rangle_{\alpha}\) represents a cluster product averaged over a collection of symmetrically inequivalent clusters labeled by the index \(\alpha\) with \(i\) being the site index and \((\varphi_{\alpha,i})_\alpha\) being a basis of orthogonal functions of the site-dependent occupation \(\sigma_i\). Multiplicity factors (\(m_{\alpha}\)) quantify how many times a symmetrically equivalent cluster appears throughout the lattice and \(J_\alpha\) is the effective cluster interaction (ECI) corresponding to the energy contribution of a cluster to the total energy.

To derive a cluster expansion, it is necessary to determine the ECIs. This process involves acquiring reference data, typically in the form of a set of configurations, along with an associated vector of target energies, which is usually obtained from first-principles calculations. Equation \eqref{eq:ce} can be expressed in a simplified vectorial form as \citep{aangqvist2019icet}
\begin{equation}
 {\boldsymbol \Delta E}= {\boldsymbol \Pi} {\boldsymbol J},
 \label{eq:ce_2}
\end{equation}
where the vector \( {\boldsymbol \Delta E}\) encodes the energies of the configurations, ${\boldsymbol J}$ represents the ECIs, and ${\boldsymbol \Pi}$ is the matrix of cluster products. The ECIs can be estimated as
\begin{equation}
{\boldsymbol J} = {\boldsymbol \Pi}^+ {\boldsymbol \Delta E}.
\label{eq:j_parameters}
\end{equation}
where ${\boldsymbol \Pi}^+ \equiv ({\boldsymbol \Pi}^\top {\boldsymbol \Pi})^{-1} {\boldsymbol \Pi}^\top $ denotes the pseudoinverse of ${\boldsymbol \Pi}$.

\subsection{Bayesian sampling}

The Bayesian approach consists of specifying a prior distribution over hypotheses or parameters. Using Bayes' theorem, as new data becomes available, the prior is combined with the likelihood to compute the posterior distribution. Implicitly, Bayes' theorem can be expressed as \citep{rasmussen2006gaussian}
\begin{equation}
\big({\sf posterior}\big) = {{\sf \big(likelihood\big)} \cdot {\sf \big(prior\big)}}/{{\sf \big(marginal\; likelihood\big)}}.
\label{eq:bayes}
\end{equation}
This approach not only enables for parameter estimation but also offers the ability to account for uncertainty and incorporate domain/empirical knowledge using Gaussian statistical distributions \(\mathcal{G}\) \citep{gelman1995bayesian}. An example of energy distribution for a collection of $N$ configurations is shown in Fig.~\ref{gaussian}. The conventional functional representation of the distribution is illustrated in Fig.~\ref{gaussian}(a). An equivalent $N$-dimensional vectorial description is shown in Fig.~\ref{gaussian}(b). The goal of the Bayesian sampling is to minimize the number of first-principles calculations by identifying a subset of configurations from a larger pool, which most effectively capture the energy trends (the energy covariance). 

\vspace{0.1cm}

\noindent\fbox{\begin{minipage}{0.97\columnwidth}
\vspace{0.05cm}
The distinct advantage of the proposed method is that the kernel of the prior statistical distribution, which encodes the energy covariance, is directly derived from  universal machine-learning potentials and physics-based potentials, rather than being modeled using a chosen metric of structural similarity.\vspace{0.05cm}
\end{minipage}}

\vspace{0.1cm}

\pagebreak

\begin{strip}
The initial step of the sampling consists of generating the prior Gaussian distribution
\begin{equation}
\mathcal{G}( {\boldsymbol \Delta E})=\left|2\pi \boldsymbol{A}^{-1}\right|^{-\frac 12}\exp\left(-\frac{1}{2}( {\boldsymbol \Delta E}-\boldsymbol{\mu})^\top \boldsymbol{A} ( {\boldsymbol \Delta E}-\boldsymbol{\mu})\right),
\label{eq:g(f)}
\end{equation}
where \(\boldsymbol \Delta E\) is a $N$-dimensional vector representing the energies of the $N$ configurations, \(\boldsymbol \mu = \EX [{\boldsymbol \Delta E}]\) is the expectation value of \(\boldsymbol \Delta E\), \(\boldsymbol A = \boldsymbol K^{-1}\) is the inverse of the covariance matrix \(\boldsymbol K\), which describes the correlations between the energies (\(\boldsymbol K=\EX [({\boldsymbol \Delta E} - \boldsymbol \mu)^\top({\boldsymbol \Delta E}-\boldsymbol \mu)]\)), and $|\cdot|$ denotes the determinant. To describe the sampling method, we rewrite \(\mathcal{G}(\boldsymbol \Delta E)\) as
\begin{equation}
\mathcal{G}({\boldsymbol \Delta E})  =  {\left|2\pi \boldsymbol A^{-1}\right|^{-\frac 12}} {\exp{\left(-\displaystyle\frac{1}{2}\begin{pmatrix}
{\boldsymbol \Delta E}_\perp-{\boldsymbol \mu}_\perp\\
{\boldsymbol \Delta E}_\parallel-{\boldsymbol \mu}_\parallel
\end{pmatrix}^\top\begin{pmatrix}
\boldsymbol{A}_{\perp\perp} & \boldsymbol{A}_{\perp\parallel}\\
\boldsymbol A_{\parallel\perp} & \boldsymbol{A}_{\parallel\parallel}
\end{pmatrix} \begin{pmatrix}
{\boldsymbol \Delta E}_\perp-{\boldsymbol \mu}_\perp\\
{\boldsymbol \Delta E}_\parallel-{\boldsymbol \mu}_\parallel
\end{pmatrix}\right)}},
\label{eq:g(f)_2}
\end{equation}
where \(\parallel \) indicates the projection on the subspace of the sampled configuration and \(\perp \) indicates the projection out of this subspace. The prior can then be refined by Bayesian inference using the configurations that have been sampled at the previous iterations. Using this information, Eq.~\eqref{eq:bayes} can be rewritten as
\begin{equation}
\mathcal{G}({\boldsymbol \Delta E}_\perp|{\boldsymbol \Delta E}_\parallel={\boldsymbol \Delta E}_0) = \left({\displaystyle\int_\perp{\mathcal{G}({\boldsymbol \Delta E}_\perp, {\boldsymbol \Delta E}_0)d{\boldsymbol \Delta E}_\perp}}\right)^{-1} {\mathcal{G}({\boldsymbol \Delta E}_\perp, {\boldsymbol \Delta E}_0)},
\label{eq:g(f)_perp}
\end{equation}
where \(\mathcal{G}({\boldsymbol \Delta E}_\perp|{\boldsymbol \Delta E}_\parallel={\boldsymbol \Delta E}_0)\) denotes the posterior distribution obtained by replacing \({\boldsymbol \Delta E}_\parallel\) with \({\boldsymbol \Delta E}_0\), which represents the energies of the already sampled configurations, \(\int_\perp{\mathcal{G}({\boldsymbol \Delta E}_\perp, {\boldsymbol \Delta E}_0)d{\boldsymbol \Delta E}_\perp}\) is the marginal likelihood and \(\mathcal{G}({\boldsymbol \Delta E}_\perp, {\boldsymbol \Delta E}_0)\) is the likelihood-weighted prior. Equation \eqref{eq:g(f)_perp} yields
\begin{equation}
\mathcal{G}({\boldsymbol \Delta E}_\perp|{\boldsymbol \Delta E}_\parallel={\boldsymbol \Delta E}_0) = \left|2\pi \boldsymbol{A}_{\perp \perp}^{-1}\right|^{-\frac 12}\exp\bigg{(}-\frac{1}{2}\big[{\boldsymbol \Delta E}_\perp -{\boldsymbol \mu}_\perp -\boldsymbol{A}_{\perp\perp}^{-1}\boldsymbol{A}_{\perp\parallel} ({\boldsymbol \Delta E}_0-{\boldsymbol \mu}_\parallel)\big]^\top \boldsymbol{A}_{\perp\perp} \big[{\boldsymbol \Delta E}_\perp-{\boldsymbol \mu}_\perp- \boldsymbol{A}_{\perp\perp}^{-1}\boldsymbol{A}_{\perp\parallel}({\boldsymbol \Delta E}_0 -{\boldsymbol \mu}_\parallel)\big]\bigg),
\label{eq:g(f)_perp2}
\end{equation}
which can be further simplified into
\begin{equation}
\mathcal{G}({\boldsymbol \Delta E}_\perp|{\boldsymbol \Delta E}_\parallel={\boldsymbol \Delta E}_0) =\left|2\pi {\boldsymbol K}_{\perp\perp}'\right|^{-\frac 12} \exp\left(-\frac{1}{2}({\boldsymbol \Delta E}_\perp-{\boldsymbol \mu}_\perp')^\top ({\boldsymbol K}_{\perp\perp}')^{-1} ({\boldsymbol \Delta E}_\perp-{\boldsymbol \mu}_\perp')\right),
\label{eq:g(f)_perp3}
\end{equation}
where ${\boldsymbol \mu}_\perp'$ and ${\boldsymbol K}_{\perp\perp}'$ stand for the mean and the covariance of the posterior, respectively. Thus, by comparing Eqs.~\eqref{eq:g(f)_perp2} and \eqref{eq:g(f)_perp3}, the mean and the covariance of the posterior can be determined iteratively as
\begin{equation}
{\boldsymbol K}_{n+1} = \left[({\boldsymbol K}_n^{-1})_{\perp \perp}\right]^{-1} 
\label{eq:k_iteration}
\end{equation}
\begin{equation}
{\boldsymbol \mu}_{n+1} = ({\boldsymbol \mu}_n)_\perp - {\boldsymbol K}_{n+1}({\boldsymbol K}_n^{-1})_{\perp \parallel}(E_{\parallel} {\boldsymbol \sigma}_\parallel-({\boldsymbol \mu}_n)_\parallel),
\label{eq:mu_iteration}
\end{equation}
\end{strip}

\begin{figure*}[ht]
    \centering
    \includegraphics[width=0.85\textwidth]{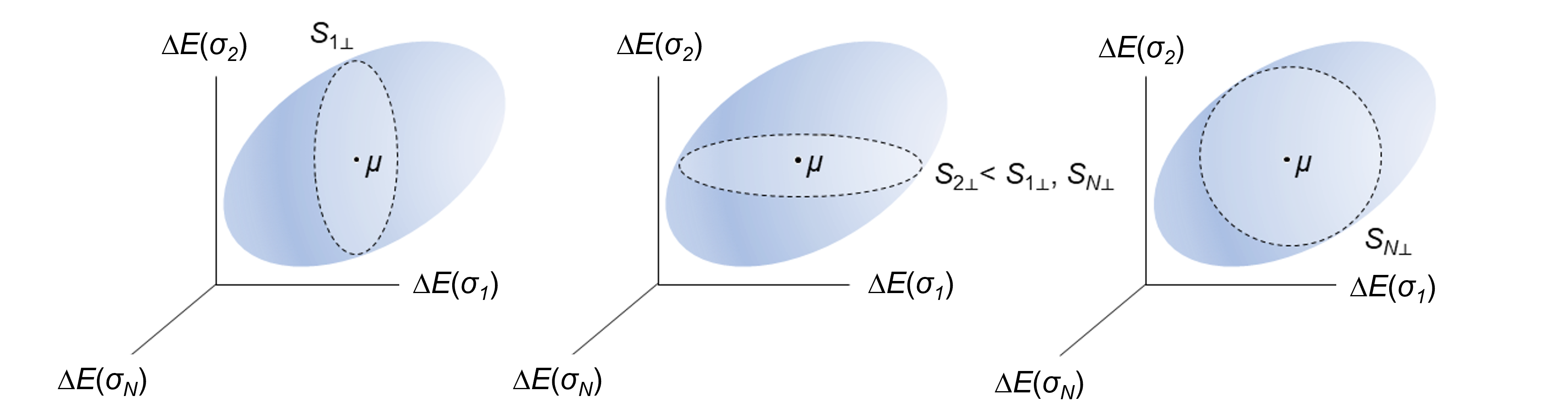}
    \caption{Evaluation of the posterior uncertainty by examining cross sections (representing the marginal probability distributions) of the prior Gaussian distribution. The estimated uncertainty of the posterior obtained by sampling configuration $\boldsymbol{\sigma}_n$ equals the area $S_{n\perp}$ of the associated cross section going through the mean $\boldsymbol \mu$ of the Gaussian distribution. The configuration to be sampled is the one minimizing the cross-sectional area (the marginal uncertainty); here, the configuration $\boldsymbol{\sigma}_2$.}
    \label{cross-section}
\end{figure*}

\noindent where \({\boldsymbol K}_{n}\) and \({\boldsymbol \mu}_{n}\) are the covariance matrix and mean vector of the prior after $n$ iterations, \(E_{\parallel}\) is the energy of the configuration ${\boldsymbol \sigma}_\parallel$ that has been newly sampled at the current ($n^{\rm th}$) iteration.

In Eqs.~\eqref{eq:k_iteration} and \eqref{eq:mu_iteration}, the configuration ${\boldsymbol \sigma}_\parallel$ that is sampled (that is, the configuration whose energy will be calculated at the DFT level at the next step of the iterative process) is the one that results in the largest reduction in the uncertainty of the posterior, which is represented as the area $S_{n\perp}$ of the $\boldsymbol \mu$-centered projection of the prior along the direction of the configuration ${\boldsymbol \sigma}_n$, as illustrated in Fig.~\ref{cross-section}. In this example, because $S_{2\perp}$ is the lowest cross-section area, the configuration ${\boldsymbol \sigma}_\parallel={\boldsymbol \sigma}_2$ will be selected, as this choice will result in maximal reduction of the posterior uncertainty. Analytically, $S_{n\perp}$ is calculated as the determinant of the covariance matrix ${\boldsymbol K}_{\perp \perp}={\boldsymbol A}^{-1}_{\perp \perp}$ after removing the row and column corresponding to that configuration from the inverse covariance matrix ${\boldsymbol A}$. One of the benefits of the Bayesian approach is the ability to quantify uncertainties $\Delta E({\boldsymbol \sigma})=2\left({\boldsymbol \sigma}^\top \boldsymbol K_{n+1} {\boldsymbol \sigma}\right)^{\frac 12}
\label{eq:std_f}$ and $\Delta J({\boldsymbol \alpha})=2\left({\boldsymbol \alpha}^\top \boldsymbol K_{n+1} {\boldsymbol \alpha}\right)^{\frac 12}$ associated to energy predictions and cluster-expansion parameters, where \(\boldsymbol \alpha = (\pi_\alpha ({\boldsymbol \sigma}_n))_n\) is the vector representing cluster $\alpha$ across the configurational space. The detailed implementation of the Bayesian sampling approach is described in the supplementary information (SI). The next section (Sec.~\ref{sec:results}) presents its application and validation in predicting the stability of Pt:Ni binaries.

\subsection{Simulations}

The Quantum ESPRESSO suite for plane-wave materials simulations was used to perform the DFT calculations \citep{giannozzi2009quantum, giannozzi2017advanced}. Projector-augmented-wave pseudopotentials from  the library \linebreak {\sc PseudoDojo} were selected to represent the ionic cores \citep{jollet2014generation} and the Perdew--Burke--Ernzerhof (PBE) \citep{perdew1996generalized} exchange-correlation functional was used to calculate the energies. The kinetic energy cutoffs for the plane waves expansion of wavefunctions and electronic charge density were set to 80 Ry and 320 Ry, respectively. To sample the Brillouin zone in reciprocal space, the ${\boldsymbol k}$-point density was set to 0.025 \AA$^{-1}$. Electronic occupations were smoothened using the Marzari--Vanderbilt cold smearing \citep{marzari1999thermal}, with a smearing width of 0.01 Ry. These kinetic energy cutoffs, ${\boldsymbol k}$-points, and smearing width were found to be sufficient to converge the total energies within 1 meV per atom and the forces within 1 meV/\AA. Classical simulations were performed in the {\sc LAMMPS} (large-scale atomic/molecular massively parallel simulator) software program \citep{Plimpton1995FastDynamics}. The interatomic potentials follow the parameterization described in Refs. \citep{Foiles1986Embedded-atom-methodAlloys, Kim2017SecondSystems, Shin2016DevelopmentCatalyst, Martinez2016PotentialPOSMat}.

\section{Results and discussion}

\label{sec:results}

To validate the Bayesian sampling, a database of 413 symmetrically unique configurations of Pt:Ni mixtures was produced, corresponding to all distinct supercells with up to eight atoms using the {\sc ICET} software package \citep{aangqvist2019icet, hart2008algorithm}. As previously stated, our objective is to decrease the number of first-principles calculations required within an extensive training set. This approach utilizes Bayesian analysis to obtain a prior from MLPs and PBPs, facilitating the recognition of the most relevant configurations in the training set. In specific terms, we employed a total of eight interatomic potentials to compute the energy of all structures in our dataset. These encompassed embedded-atom potentials (EAM \citep{Foiles1986Embedded-atom-methodAlloys}, MEAM \citep{Kim2017SecondSystems}), reactive many-body potentials (COMB3 \citep{liang2013classical}, ReaxFF \citep{Shin2016DevelopmentCatalyst}), and machine-learned potentials (CHGNet \citep{deng2023chgnet}, M3GNet \citep{chen2022universal}, ALIGNN \citep{choudhary2021atomistic}, MACE \citep{Batatia2022mace,Batatia2022Design}).

\begin{figure}[ht]
    \centering
    \includegraphics[width=\columnwidth]{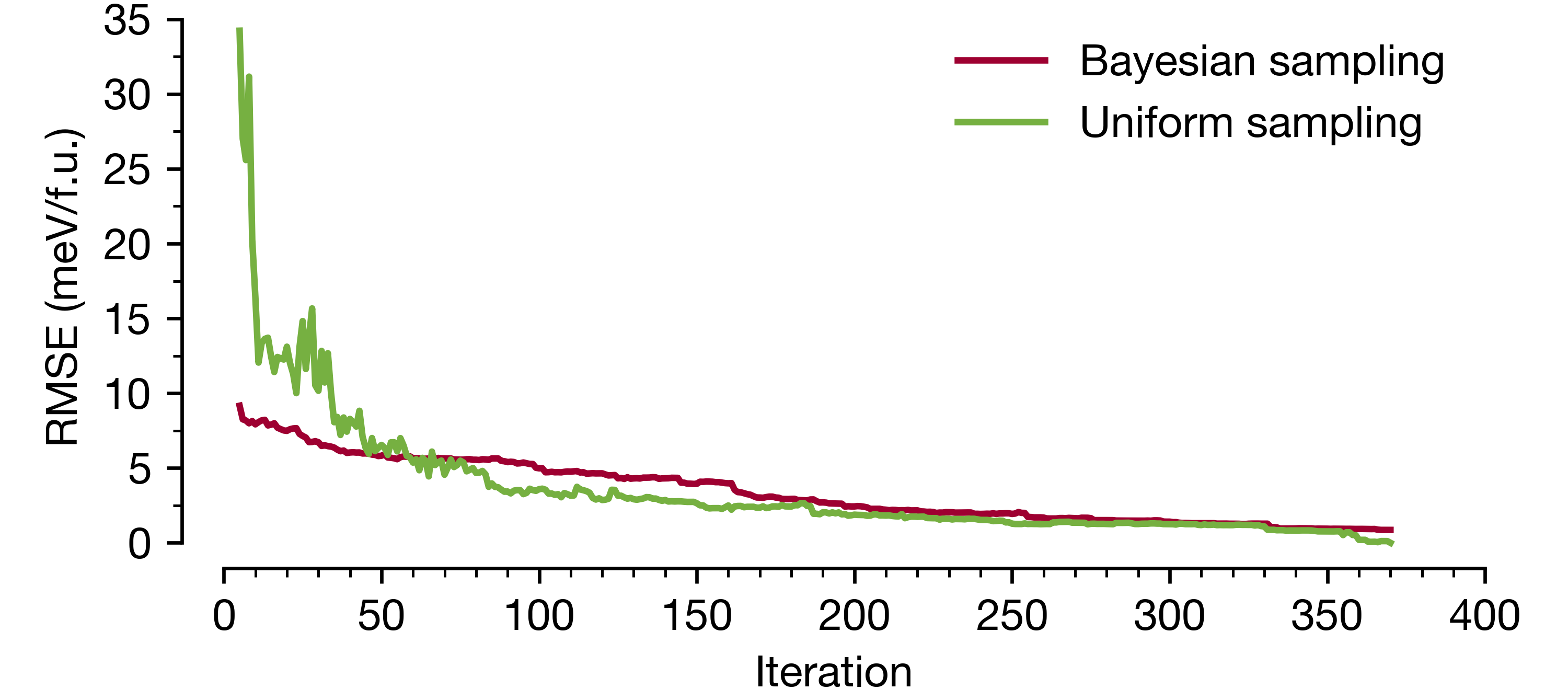}
    \caption{Root mean squared error (RMSE) of the cluster expansion using Bayesian and uniform sampling as a function of the number of sampling iterations (the number of DFT calculations). RMSE is estimated against a cluster expansion derived from the complete dataset of fully converged DFT energies.}
    \label{rmse}
\end{figure}

Our approach relies on the premise that the prior captures correlations between the formation energies of the configurations. We systematically tested this hypothesis by comparing the formation energies calculated from DFT to those computed with the interatomic potentials mentioned in the preceding paragraph. As shown in Fig.~\ref{ml parity}(a), there exist significant discrepancies between the DFT and MLP/PBP energies. However, these discrepancies do not imply that these potentials cannot provide exploitable information. In fact, upon renormalizing the empirical energies from MLPs or PBPs by the scaling factor
\begin{equation}
\alpha= \left[{(\boldsymbol \Delta \tilde E)^\top\boldsymbol \Delta \tilde E}\right]^{-1} \left[{(\boldsymbol \Delta \tilde E)}^\top \boldsymbol \Delta E \right]
\label{eq:scale}
\end{equation}
(where \(\boldsymbol \Delta E\) represents the DFT energies and \(\boldsymbol \Delta \tilde E\) is the energy calculated using the MLP/PBP empirical potential), a close correspondence is found between the DFT and empirical trends, suggesting that the ordering of the calculated empirical energies is qualitatively consistent with its DFT counterpart [Fig.~\ref{ml parity}(b)]. In practice, the calculation of the rescaling factor is repeated for all the interatomic potentials at each iteration, allowing for the gradual improvement of the empirical trends with the progressive incorporation of new DFT energies. After few iterations, the rescaled potentials closely capture energy correlations.

\begin{figure*}[ht]
    \centering
    \includegraphics[width=\textwidth]{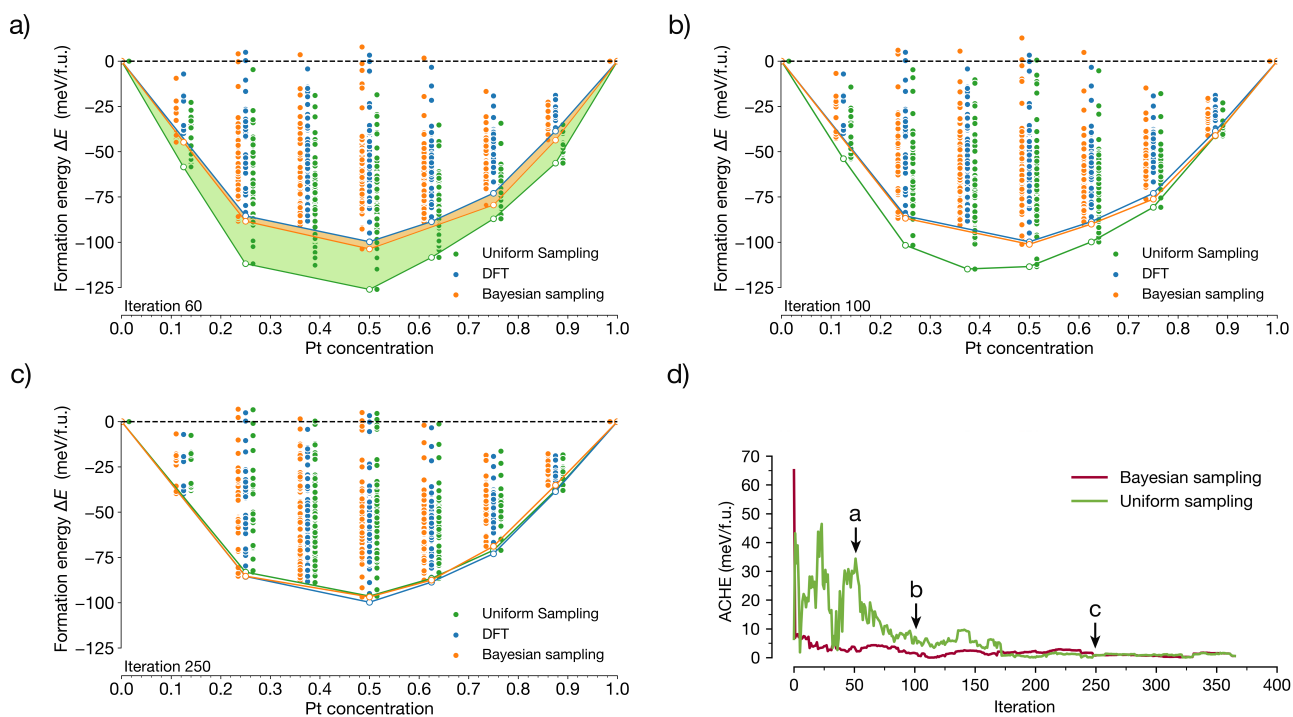}
    \caption{Cluster expansion performance as a function of the number of sampling iterations. The energy points associated with uniform (Bayesian) sampling are slightly shifted to the right (left) of the actual concentrations for ease of comparison. Convex hull diagrams after (a) 60 iterations, (b) 100 iterations, and (c) 250 iterations. (d) Areal convex hull error (ACHE) between the Bayesian sampling and DFT, and uniform sampling and DFT as function of the number of iterations. The areal convex hull errors (ACHEs) are shown as the colored (orange and green) transparent regions in panel (a) for Bayesian and uniform sampling, respectively.}
    \label{area}
\end{figure*}

Next, a cluster expansion was parameterized, incorporating clusters up to the fourth order, with cutoff distances of 10 \AA\ for pairs, 7.5 \AA\ for triplets, and 5 \AA\ for quadruplets. This cluster space was composed of a total of 130 parameters, distributed as follows: 1 zerolet, 1 singlet, 17 pairs, 76 triplets, and 35 quadruplets. To analyze how the performance of the approach is affected by increasing the number of DFT calculations in each iteration, we generated a learning curve by assessing the root mean squared error (RMSE) against a cluster expansion derived solely from DFT calculations. It should be mentioned that 370 out of the initial 413 structures were successfully converged during the DFT calculations. Structures that did not converge were excluded from our interatomic potentials database. 

We derived the prior distribution by utilizing eight different interatomic potentials with statistical weights representing the amount of information contained in each of them, as explained in Sec.~S4 of the SI. The prior was employed in the iterative process of minimizing uncertainty to select the optimal configurations. The performance of the resulting Bayesian sampling method is compared to randomly sampling (uniform sampling) the structures in Fig.~\ref{rmse}, which depicts the root mean squared error of the Bayesian and random cluster expansion models with respect to a cluster expansion model solely derived from DFT calculations. A notable difference in the convergence of the RMSE is observed, especially at the initial stages of iterations where Bayesian sampling leads to an immediate decrease of the RMSE. After the 60$^{\rm th}$ iteration, uniform sampling seems to outperform Bayesian sampling because it may better capture configurations away from the convex hull where the accuracy of MLPs/PBPs is expected to deteriorate (as these high-energy configurations are generally less represented in MLP/PBP training). However, as shown in Fig.~\ref{area}(a), the convex hull generated using uniform sampling at iteration 60 noticeably differs from the convex hull created using all available training structures, while the Bayesian hull is already very close to the DFT target.

To correctly assess the convergence of the convex hull, we introduce a direct metric of convex hull accuracy, the areal convex hull error (ACHE), obtained by calculating the area between the convex hulls, as depicted in Fig.~\ref{area}. Changes in ACHE along the iterative cycle are reported in Fig.~\ref{area}(d). A noteworthy observation is the close alignment between the Bayesian and DFT curves after 20-40 iterations, while uniform sampling requires 150-170 iterations to reach an ACHE accuracy of 3 meV. Additionally, with Bayesian sampling, the correct prediction of the convex hull is achieved after 100 iterations, whereas uniform sampling provides consistent predictions only after 250 iterations. These observations demonstrate that Bayesian sampling significantly reduces the number of DFT steps in building well-converged cluster expansions.

It is worth noting that further computational acceleration would be achieved by opting for a batch selection strategy at each iteration (as opposed to processing individual structures) and by conducting parallel DFT calculations for the selected batch. To assess the effectiveness of this approach, we conducted a test by calculating the DFT energy for 5 or 10 structures in each iteration. The results demonstrated a marginal increase in the number of DFT calculations required. With a batch of 5, the model achieved convergence after 105 DFT calculations, while for a batch of 10, convergence was attained after 110 DFT calculations. In contrast, the single-structure selection approach reached convergence after 100 DFT calculations. Therefore, it is advisable to employ batch selection to minimize computational time in generating accurate cluster expansions.

\section{Conclusion}

We introduced a Bayesian selection algorithm to expedite the robust parameterization of accurate cluster expansions using covariance information extracted from machine-learned and physics-based interatomic potentials. This prior enables one to identify the most informative structures within a training set for model construction. The energies of the selected structures are calculated at the DFT level. The prior is then updated by incorporating the computed DFT energies. Applying this iterative approach to a prototypical Pt:Ni alloy provided well-converged CE at a fraction of the computational cost of uniform sampling. Importantly, much lower statistical fluctuations were observed using Bayesian inference. Further acceleration was attained by selecting a batch of structures at each iteration rather than performing DFT calculations for single structures. This algorithm provides a powerful approach for future studies of multicomponent interfaces and materials at finite temperature.

\section{Acknowledgments}

This work was primarily supported by the U.S. Department of Energy, Office of Science, Basic Energy Sciences, CPIMS (Condensed Phase and Interfacial Molecular Science) Program, under Award No.~DE-SC0018646. S.G.~acknowledges support from the Center for Nanoscale Science under Grant No.~DMR-2011839 (code implementation, code validation, and development of accuracy metrics). I.D.~and A.D.~are thankful to Paul E. Lammert for fruitful discussions on the analytical foundations of the cluster expansion method.

\bibliography{bibliography.bib}

\begin{thebibliography}{10}

\bibitem{liang2013classical}
Tao Liang, Tzu-Ray Shan, Yu-Ting Cheng, Bryce~D Devine, Mark Noordhoek, Yangzhong Li, Zhize Lu, Simon~R Phillpot, and Susan~B Sinnott.
\newblock Classical atomistic simulations of surfaces and heterogeneous interfaces with the charge-optimized many body (comb) potentials.
\newblock {\em Materials Science and Engineering: R: Reports}, 74(9):255--279, 2013.

\bibitem{Shin2016DevelopmentCatalyst}
Yun~Kyung Shin, Lili Gai, Sumathy Raman, and Adri~C.T. Van~Duin.
\newblock {Development of a ReaxFF Reactive Force Field for the Pt-Ni Alloy Catalyst}.
\newblock {\em Journal of Physical Chemistry A}, 120(41):8044--8055, 10 2016.

\bibitem{Foiles1986Embedded-atom-methodAlloys}
S~M Foiles, M~I Baskes, and M~S Daw.
\newblock {Embedded-atom-method functions for the fcc metals Cu, Ag, Au, Ni, Pd, Pt, and their alloys}.
\newblock {\em Physical Review B}, 33(12):7983--7991, 6 1986.

\bibitem{Kim2017SecondSystems}
Jin-Soo Kim, Donghyuk Seol, Joonho Ji, Hyo-Sun Jang, Yongmin Kim, and Byeong-Joo Lee.
\newblock {Second nearest-neighbor modified embedded-atom method interatomic potentials for the Pt-M (M = Al, Co, Cu, Mo, Ni, Ti, V) binary systems}.
\newblock {\em Calphad}, 59:131--141, 12 2017.

\bibitem{deng2023chgnet}
Bowen Deng, Peichen Zhong, KyuJung Jun, Janosh Riebesell, Kevin Han, Christopher~J Bartel, and Gerbrand Ceder.
\newblock Chgnet as a pretrained universal neural network potential for charge-informed atomistic modelling.
\newblock {\em Nature Machine Intelligence}, 5(9):1031--1041, 2023.

\bibitem{chen2022universal}
Chi Chen and Shyue~Ping Ong.
\newblock A universal graph deep learning interatomic potential for the periodic table.
\newblock {\em Nature Computational Science}, 2(11):718--728, 2022.

\bibitem{choudhary2021atomistic}
Kamal Choudhary and Brian DeCost.
\newblock Atomistic line graph neural network for improved materials property predictions.
\newblock {\em npj Computational Materials}, 7(1):185, 2021.

\bibitem{Batatia2022mace}
Ilyes Batatia, David~P Kovacs, Gregor Simm, Christoph Ortner, and G{\'a}bor Cs{\'a}nyi.
\newblock Mace: Higher order equivariant message passing neural networks for fast and accurate force fields.
\newblock {\em Advances in Neural Information Processing Systems}, 35:11423--11436, 2022.

\bibitem{Batatia2022Design}
Ilyes Batatia, Simon Batzner, D{\'a}vid~P{\'e}ter Kov{\'a}cs, Albert Musaelian, Gregor N.~C. Simm, Ralf Drautz, Christoph Ortner, Boris Kozinsky, and G{\'a}bor Cs{\'a}nyi.
\newblock The design space of e(3)-equivariant atom-centered interatomic potentials, 2022.

\bibitem{eckhoff2023lifelong}
Marco Eckhoff and Markus Reiher.
\newblock Lifelong machine learning potentials.
\newblock {\em Journal of Chemical Theory and Computation}, 2023.

\bibitem{kadkhodaei2021cluster}
Sara Kadkhodaei and Jorge~A Mu{\~n}oz.
\newblock {Cluster expansion of alloy theory: a review of historical development and modern innovations}.
\newblock {\em JOM}, 73(11):3326--3346, 2021.

\bibitem{nelson2013cluster}
{Nelson, Lance J and Ozoli{\c{n}}{\v{s}}, Vidvuds and Reese, C Shane and Zhou, Fei and Hart, Gus LW}.
\newblock {Cluster expansion made easy with Bayesian compressive sensing}.
\newblock {\em Physical Review B}, 88(15):155105, 2013.

\bibitem{wolverton1998prediction}
C~Wolverton and Alex Zunger.
\newblock {Prediction of Li Intercalation and Battery Voltages in Layered vs. Cubic Li$_x$CoO2}.
\newblock {\em Journal of The Electrochemical Society}, 145(7):2424, 1998.

\bibitem{seko2006prediction}
Atsuto Seko, Koretaka Yuge, Fumiyasu Oba, Akihide Kuwabara, and Isao Tanaka.
\newblock {Prediction of ground-state structures and order-disorder phase transitions in II-III spinel oxides: A combined cluster-expansion method and first-principles study}.
\newblock {\em Physical Review B}, 73(18):184117, 2006.

\bibitem{kolb2005nonmetal}
Brian Kolb and Gus~LW Hart.
\newblock {Nonmetal ordering in TiC1-xNx: Ground-state structure and the effects of finite temperature}.
\newblock {\em Physical Review B}, 72(22):224207, 2005.

\bibitem{carlsson2023finding}
Adam Carlsson, Johanna Rosen, and Martin Dahlqvist.
\newblock {Finding stable multi-component materials by combining cluster expansion and crystal structure predictions}.
\newblock {\em npj Computational Materials}, 9(1):21, 2023.

\bibitem{wu2016cluster}
Qu~Wu, Bing He, Tao Song, Jian Gao, and Siqi Shi.
\newblock {Cluster expansion method and its application in computational materials science}.
\newblock {\em Computational Materials Science}, 125:243--254, 2016.

\bibitem{ozolicnvs1998cu}
V~Ozoli{\c{n}}{\v{s}}, C~Wolverton, and Alex Zunger.
\newblock {Cu-Au, Ag-Au, Cu-Ag, and Ni-Au intermetallics: First-principles study of temperature-composition phase diagrams and structures}.
\newblock {\em Physical Review B}, 57(11):6427, 1998.

\bibitem{sanchez1989magnetic}
JM~Sanchez, JL~Moran-Lopez, C~Leroux, and MC~Cadeville.
\newblock {Magnetic properties and chemical ordering in Co-Pt}.
\newblock {\em Journal of Physics: Condensed Matter}, 1(2):491, 1989.

\bibitem{asta1993theoretical}
Mark Asta, Ryan McCormack, and Didier de~Fontaine.
\newblock {Theoretical study of alloy phase stability in the Cd-Mg system}.
\newblock {\em Physical Review B}, 48(2):748, 1993.

\bibitem{van2005vacancies}
A~Van~der Ven and G~Ceder.
\newblock {Vacancies in ordered and disordered binary alloys treated with the cluster expansion}.
\newblock {\em Physical Review B}, 71(5):054102, 2005.

\bibitem{cao2018use}
Liang Cao, Chenyang Li, and Tim Mueller.
\newblock {The use of cluster expansions to predict the structures and properties of surfaces and nanostructured materials}.
\newblock {\em Journal of Chemical Information and Modeling}, 58(12):2401--2413, 2018.

\bibitem{van2002automating}
Axel van~de Walle and Gerbrand Ceder.
\newblock {Automating first-principles phase diagram calculations}.
\newblock {\em Journal of Phase Equilibria}, 23(4):348, 2002.

\bibitem{seko2009cluster}
Atsuto Seko, Yukinori Koyama, and Isao Tanaka.
\newblock {Cluster expansion method for multicomponent systems based on optimal selection of structures for density-functional theory calculations}.
\newblock {\em Physical Review B}, 80(16):165122, 2009.

\bibitem{mueller2010exact}
Tim Mueller and Gerbrand Ceder.
\newblock {Exact expressions for structure selection in cluster expansions}.
\newblock {\em Physical Review B}, 82(18):184107, 2010.

\bibitem{cockayne2010building}
Eric Cockayne and Axel van~de Walle.
\newblock {Building effective models from sparse but precise data: Application to an alloy cluster expansion model}.
\newblock {\em Physical Review B}, 81(1):012104, 2010.

\bibitem{mueller2009bayesian}
Tim Mueller and Gerbrand Ceder.
\newblock {Bayesian approach to cluster expansions}.
\newblock {\em Physical Review B}, 80(2):024103, 2009.

\bibitem{nelson2013compressive}
Lance~J Nelson, Gus~LW Hart, Fei Zhou, Vidvuds Ozoli{\c{n}}{\v{s}}, et~al.
\newblock {Compressive sensing as a paradigm for building physics models}.
\newblock {\em Physical Review B}, 87(3):035125, 2013.

\bibitem{tibshirani1996regression}
Robert Tibshirani.
\newblock {Regression shrinkage and selection via the lasso}.
\newblock {\em Journal of the Royal Statistical Society Series B: Statistical Methodology}, 58(1):267--288, 1996.

\bibitem{packwood2017bayesian}
Daniel Packwood et~al.
\newblock {\em {Bayesian Optimization for Materials Science}}.
\newblock Springer, 2017.

\bibitem{todorovic2019bayesian}
Milica Todorovi{\'c}, Michael~U Gutmann, Jukka Corander, and Patrick Rinke.
\newblock {Bayesian inference of atomistic structure in functional materials}.
\newblock {\em npj Computational Materials}, 5(1):35, 2019.

\bibitem{sanchez2016approximate}
JM~Sanchez and T~Mohri.
\newblock {Approximate solutions to the cluster variation free energies by the variable basis cluster expansion}.
\newblock {\em Computational Materials Science}, 122:301--306, 2016.

\bibitem{aangqvist2019icet}
Mattias {\AA}ngqvist, William~A Mu{\~n}oz, J~Magnus Rahm, Erik Fransson, C{\'e}line Durniak, Piotr Rozyczko, Thomas~H Rod, and Paul Erhart.
\newblock {ICET--a Python library for constructing and sampling alloy cluster expansions}.
\newblock {\em Advanced Theory and Simulations}, 2(7):1900015, 2019.

\bibitem{rasmussen2006gaussian}
Carl~Edward Rasmussen and Christopher~KI Williams.
\newblock {\em {Gaussian Processes for Machine Learning}}, volume~1.
\newblock Springer, 2006.

\bibitem{gelman1995bayesian}
Andrew Gelman, John~B Carlin, Hal~S Stern, and Donald~B Rubin.
\newblock {\em {Bayesian Data Analysis}}.
\newblock Chapman and Hall/CRC, 1995.

\bibitem{giannozzi2009quantum}
Paolo Giannozzi, Stefano Baroni, Nicola Bonini, Matteo Calandra, Roberto Car, Carlo Cavazzoni, Davide Ceresoli, Guido~L Chiarotti, Matteo Cococcioni, Ismaila Dabo, et~al.
\newblock {QUANTUM ESPRESSO: a modular and open-source software project for quantum simulations of materials}.
\newblock {\em Journal of Physics: Condensed Matter}, 21(39):395502, 2009.

\bibitem{giannozzi2017advanced}
Paolo Giannozzi, Oliviero Andreussi, Thomas Brumme, Oana Bunau, M~Buongiorno Nardelli, Matteo Calandra, Roberto Car, Carlo Cavazzoni, Davide Ceresoli, Matteo Cococcioni, et~al.
\newblock {Advanced capabilities for materials modelling with Quantum ESPRESSO}.
\newblock {\em Journal of Physics: Condensed Matter}, 29(46):465901, 2017.

\bibitem{jollet2014generation}
Fran{\c{c}}ois Jollet, Marc Torrent, and Natalie Holzwarth.
\newblock {Generation of Projector Augmented-Wave atomic data: A 71 element validated table in the XML format}.
\newblock {\em Computer Physics Communications}, 185(4):1246--1254, 2014.

\bibitem{perdew1996generalized}
John~P Perdew, Kieron Burke, and Matthias Ernzerhof.
\newblock {Generalized gradient approximation made simple}.
\newblock {\em Physical Review Letters}, 77(18):3865, 1996.

\bibitem{marzari1999thermal}
Nicola Marzari, David Vanderbilt, Alessandro De~Vita, and MC~Payne.
\newblock {Thermal contraction and disordering of the Al (110) surface}.
\newblock {\em Physical Review Letters}, 82(16):3296, 1999.

\bibitem{Plimpton1995FastDynamics}
Steve Plimpton.
\newblock {Fast Parallel Algorithms for Short-Range Molecular Dynamics}.
\newblock {\em Journal of Computational Physics}, 117(1):1--19, 3 1995.

\bibitem{Martinez2016PotentialPOSMat}
Jackelyn~A. Martinez, Aleksandr Chernatynskiy, Dundar~E. Yilmaz, Tao Liang, Susan~B. Sinnott, and Simon~R. Phillpot.
\newblock {Potential Optimization Software for Materials (POSMat)}.
\newblock {\em Computer Physics Communications}, 203:201--211, 6 2016.

\bibitem{hart2008algorithm}
Gus~LW Hart and Rodney~W Forcade.
\newblock {Algorithm for generating derivative structures}.
\newblock {\em Physical Review B}, 77(22):224115, 2008.

\end{thebibliography}

\end{document}